\documentstyle[12pt]{article}
\begin{document}
\def\emline#1#2#3#4#5#6{%
       \put(#1,#2){\special{em:moveto}}%
       \put(#4,#5){\special{em:lineto}}}
\def\newpic#1{}

\thispagestyle{empty}
\begin{titlepage}
\begin{center}
{\LARGE\bf Heavy baryons in the relativistic quark model}

\vspace{3em}

{\large D. Ebert and R. N. Faustov \\[0.3em]
\small\it Institut f\"ur Physik, Humboldt--Universit\"at zu Berlin,\\[0.3em]
\small\it Invalidenstr.110, D-10115 Berlin}
\vspace{2em}

{\large V. O. Galkin \\[0.3em]
\small\it Russian Academy of Sciences, Scientific Council for
Cybernetics,\\[0.3em]
\small\it Vavilov Street 40, Moscow 117333, Russia}
\vspace{2em}

{\large A. P. Martynenko and V. A. Saleev \\[0.3em]
\small\it Samara State University, Samara 443011, Russia}

\end{center}
\end{titlepage}


\renewcommand{\thefootnote}{\fnsymbol{footnote}}
\title{\bf Heavy baryons in the relativistic quark model}
\author{ D. Ebert and R. N. Faustov\footnotemark[1]{}\ \footnotemark[2] \\
\small\it Institut f\"ur Physik, Humboldt--Universit\"at zu Berlin,\\
\small\it Invalidenstr.110, D-10115 Berlin\\
\\
V. O. Galkin \\
\small\it Russian Academy of Sciences, Scientific Council for
Cybernetics,\\
\small\it Vavilov Street 40, Moscow 117333, Russia\\
\\
A. P. Martynenko and V. A. Saleev \\
\small\it Samara State University, Samara 443011, Russia}
\date{}
\maketitle

\setcounter{footnote}{1}
\footnotetext{Supported by
Graduiertenkolleg {\it Elementarteilchenphysik} and
in part by {\it Russian Foundation for Fundamental Research}
 under Grant No.\ 96-02-17171.}
\setcounter{footnote}{2}
\footnotetext{On leave of absence from Russian Academy of Sciences,
Scientific Council for Cybernetics,
Vavilov Street 40, Moscow 117333, Russia.}
\renewcommand{\thefootnote}{\arabic{footnote}}
\setcounter{footnote}{0}

\begin{abstract}
In the framework of the relativistic quasipotential quark model the
mass spectrum of baryons with two heavy quarks is calculated. The
quasipotentials for interactions of two quarks and of a quark with a
scalar and axial vector diquark are evaluated. The bound state masses
of baryons with $J^P=\frac{1}{2}^+, \frac{3}{2}^+$ are computed.
\end{abstract}

\section{Introduction}
The investigation of properties of hadrons, containing heavy quarks,
is of a great interest for understanding the dynamics of quark and
gluon interaction. To calculate the meson and baryon mass spectra
QCD sum rules \cite{S}, potential models \cite{GI,Bagan,KLT,Richard},
lattice QCD \cite{Al}, heavy quark effective theory (HQET)
\cite{Savage,Korner}, the method of vacuum correlators \cite{Simon} and
some other approaches are widely used.
Presently at the LHC, $B$-factories and the Tevatron with high luminosity,
several experiments have been proposed, in which a detailed study of
baryons containing two heavy quarks can be performed. The
possible quark composition of such baryons looks as follows: ($ccq$,
$cbq$, $bbq$), where $c$ and $b$ are heavy quarks ($Q$) and $q$
denotes a light $u$, $d$ and $s$ quark. In this connection theoretical
predictions for heavy baryon masses acquire important significance.
The properties of heavy hadrons (containing $b$ and/or $c$ quarks) are
essentially different from those of light hadrons composed from $u$,
$d$ and $s$ quarks. This difference originates from the fact that
the heavy quark mass strongly exceeds the scale of QCD interaction,
$\Lambda_{QCD}\approx 300 - 400$ MeV, namely
$m_{c,b}\gg\Lambda_{QCD}$. Therefore, while investigating processes
with heavy quarks one should expect that the limit $m_Q\to\infty$
within QCD yields good quantitative results, which may be further
improved in HQET by taking into account $1/m_Q$ corrections. Baryons, containing two
heavy quarks ($QQq$), may be considered as a localized source ($QQ$) of the
colour field, in which the light quark moves. Interaction
forces between heavy quarks then lead to the formation of a two-particle
bound state of a $QQ$-diquark, the scale of which is determined by the
quantity $1/m_Q$ which is small compared to the QCD scale
$1/\Lambda_{QCD}$. Thereby it seems justified to treat the heavy
diquark as a pointlike object with definite colour, spin and mass and to ignore
the configuration with a $qQ$-diquark which size is of order $1/m_q\sim 1/\Lambda_{QCD}$.
The baryon formation then occurs as a result of a diquark interaction
with the light quark. Here in the framework of the quark-diquark
approximation we calculate the mass spectrum of heavy baryons on the
basis of a local Schr\"odinger-like quasipotential equation \cite{MF}:
\begin{equation}
\left(\frac{b^2(M)}{2\mu_R}-\frac{
\vec p^2}{2\mu_R}\right)\psi_M(\vec p)=
\int\frac{d^3q}{(2\pi)^3}V(\vec p,\vec q,M)\psi_M(\vec q),
\end{equation}
where the relativistic reduced mass is
\begin{equation}
\mu_R=\frac{E_1E_2}{E_1+E_2}=\frac{M^4-(m_1^2-m_2^2)^2}{4M^3},
\end{equation}
\begin{displaymath}
E_1=\frac{M^2-m_2^2+m_1^2}{2M},\quad E_2=\frac{M^2-m_1^2+m_2^2}{2M},\quad
E_1+E_2=M,
\end{displaymath}
and the center of mass system relative momentum squared on the mass
shell reads
\begin{equation}
b^2(M)=\frac{[M^2-(m_1+m_2)^2][M^2-(m_1-m_2)^2]}{4M^2},
\end{equation}
with $m_{1,2}$ the masses of the constituent particles. In refs.
\cite{GF,GMF} the operator of quark-antiquark interaction has been
constructed and the mass spectra and decay rates of mesons have been
investigated. The relativistic quasipotential quark model gives in the
meson sector results which nicely agree with experimental data.
Relativistic effects play an important role in describing properties
of quark bound states and may be consistently taken into account
within the quasipotential method \cite{FMS}.

\section{Diquarks in the relativistic quasipotential model}
The concept of diquarks was introduced at the very beginning of
the development of the quark model of hadrons in order to describe
baryon properties \cite{Richard}. Diquarks represent by themselves
two-particle clusters, which are produced inside the three-quark
systems as a result  of quark interactions in which the spin-dependent
forces play an important role \cite{Fleck,An}. The attractive forces
between two quarks act if the quarks are in the antisymmetric colour
state. The interaction potential at large distances in the quark-quark
system is similar to that in the quark-antiquark system since the
baryon size is practically the same as the meson size. In
the ground state the diquarks are two-particle bound states of
quarks in an antisymmetric colour state with zero angular momentum and definite
flavour and spin. In the case of indentical quarks the
diquark has spin $S=1$, whereas for quarks of different flavours the
state with spin $S=0$ is also possible.

Now we construct the interaction quasipotential for two heavy quarks
$Q$, which enters in eq.~(1). The interaction operator in the $QQ$
system contains a perturbative part, which is defined at small distances
in the quasipotential approach by the one-gluon exchange amplitude and
can be evaluated within QCD. Moreover there is a nonperturbative term
in the operator $V(\vec p,\vec q, M)$, which cannot be obtained
consistently within QCD and leads to the quark confinement at large
distances.

The quasipotential $V(\vec p,\vec q, M)$ of eq.~(1) is constructed with
the help of the off-mass-shell scattering amplitude, projected onto
the positive energy states. Then the perturbative part of the
quasipotential can be represented in the form:
\begin{equation}
V_{QQ}^{pert.}(\vec p,\vec q)=\bar u_1(\vec p)\bar u_2(-\vec p)
\frac{\frac{2}{3}\alpha_s}{4\sqrt{\epsilon_1(\vec p)\epsilon_1(\vec q)
\epsilon_2(\vec p)\epsilon_2(\vec q)}}
D_{\mu\nu}(k)\gamma_1^\mu\gamma_2^\nu u_1(\vec q)u_2(-\vec q),
\end{equation}
where $D_{\mu\nu}(k)$ is the gluon propagator conveniently taken in the Coulomb gauge
\begin{equation}
D^{00}(k)=-\frac{1}{\vec
k^2},~~~~D^{ij}(k)=-\frac{1}{k^2}\left(\delta^{ij}-\frac{k^ik^j}{\vec
k^2}\right),~~~D^{0i}=D^{i0}=0,
\end{equation}
$\vec k=\vec p-\vec q$, and
$u(\vec p)$ is the Dirac spinor,
\begin{equation}
u^\lambda(\vec p)=\sqrt{\varepsilon(\vec p)+m}
{1\choose \frac{\vec\sigma\vec p}{\varepsilon(\vec p)+m}}\chi^\lambda.
\end{equation}
In this case the colour factor, which includes the colour part of the
baryon wave function $\varepsilon_{ijk}/\sqrt{6}$ ($i, j, k=1,2,3$)
has the value:
\begin{equation}
\frac{1}{\sqrt{6}}\varepsilon_{ijk}\delta_{nk}\frac{1}{\sqrt{6}}\varepsilon_
{lmn}T^a_{li}T^a_{mj}=-\frac{2}{3},
\end{equation}
where $T^a$ are the generators of the colour $SU_c(3)$ group. It
means, that in the two-quark system the exchange of a nonabelian gauge
boson results in an attractive Coulomb potential at small distances
which differs only by a factor $1/2$ from the similar potential in the
quark-antiquark system. Substituting the spinor (6) into eq.(4) and carrying
out some manipulations with spinor convolutions, we obtain, in the
same way as in \cite{GMF}, the one-gluon exchange quasipotential
including relativistic corrections. Since we consider the ground
state of diquarks, the angular momentum of the $QQ$ system is zero. As
it has been remarked in ref.~\cite{Narod}, the quark spin interaction
can result in the existence of short-range correlations in the
quark-quark subsystem. Therefore, neglecting
corrections $O(p^2/m_Q^2)$, we can represent the perturbative part of
the quasipotential in the configuration space in the form:
\begin{equation}
V_{QQ}^{pert.}(r)=-\frac{2}{3}\frac{\alpha_s}{r}+\frac{4\pi\alpha_s}{9m_1m_2}
(\vec\sigma_1\vec\sigma_2)\delta(\vec r),
\end{equation}
where $\alpha_s(r)$ is the running coupling constant and $\sigma_{1,2}$
are the quark spin matrices. The colour hyperfine interaction is
attractive if the total spin of two quarks $S=0$, and repulsive in the
case $S=1$.

In the approximation of one-gluon exchange the quark-quark potential in a
baryon is equal to one half of the quark-antiquark potential in
a meson. Calculations based on the baryon Wilson loop area law in QCD \cite{Cor}
show that this
relation holds even true for nonperturbative
interactions. Thus, while constructing the long-range nonperturbative
part of the quark interaction operator we used the known ``$1/2$
rule", i.~e. $V_{qq}=\frac{1}{2}V_{q\bar q}$ \cite{Richard} and chose
this part of the quasipotential in the standard linearly rising in
$r$ form. As a result in this approximation the complete interaction
operator in the quark-quark system is chosen in the form:
\begin{equation}
V_{QQ}(r)=-\frac{2}{3}\frac{\alpha_s}{r}+\frac{1}{2}(Ar+B)+\frac{4\pi\alpha_s}
{9m_1m_2}(\vec\sigma_1\vec\sigma_2)\delta(\vec r),
\end{equation}
where the parameters $A=0.18$ GeV$^2$ and $B=-0.3$ GeV had been fixed
previously \cite{GF,GMF} in calculating the meson mass spectrum. The
numerical solution of the quasipotential equation (1) with interaction
operator (9) yields the mass spectrum of the $s$-wave axial vector and scalar
diquarks. The results are presented \footnote{The masses
of light diquarks are presented for completeness where $q = u,d$.} in Table ~1. Here the usual
values of quark masses (the same as for mesons) were used \cite{GF,GMF}:

$m_u=m_d=0.33$ GeV, $m_s=0.5$ GeV, $m_c=1.55$ GeV, $m_b=4.88$ GeV.
\begin{table*}[hbt]
\caption{Masses of $s$-wave scalar and axial vector diquarks}
\begin{center}
\begin{tabular}{|c|c|c|c|c|c|c|c|c|c|c|}    \hline Diquark
& qq & qs & qc  & qb  & ss  & sc & sb & cc & cb & bb  \\       \hline
M(GeV) &0.85 &1.04&2.08 & 5.39& -   &2.27 &5.57&- &6.52&   -   \\
(scalar diquark) &     &     &     &     &     &     &     &     &
& \\  \hline M(GeV)   &1.02 & 1.15  &2.12 & 5.41& 1.32& 2.29& 5.58&
3.26&6.52& 9.79 \\ (axial vector diquark)    &     &     &     &     &
&     &     & &     &     \\   \hline \end{tabular}
\end{center}
\end{table*}

\section{Quasipotential for the interaction of a quark with a scalar
diquark}
Now to calculate the baryon mass spectrum in the quark-diquark
approximation it is necessary to construct the quark-diquark
interaction operator. In the case of a scalar diquark the perturbative
part of the quasipotential is described by the Feynman diagram in
Fig.~1.

\begin{figure}[htb]
\unitlength=1.00mm
\special{em:linewidth 1pt}
\linethickness{1pt}
\begin{picture}(130.00,52.04)
\emline{40.00}{49.89}{1}{120.00}{49.89}{2}
\emline{120.00}{9.89}{3}{40.00}{9.89}{4}
\emline{40.00}{12.90}{5}{120.00}{12.90}{6}
\put(80.00,11.18){\circle*{3.44}}
\emline{80.00}{49.89}{7}{83.00}{46.88}{8}
\emline{83.00}{46.88}{9}{77.00}{43.87}{10}
\emline{77.00}{43.87}{11}{83.00}{40.86}{12}
\emline{83.00}{40.86}{13}{77.00}{37.85}{14}
\emline{77.00}{37.85}{15}{83.00}{34.84}{16}
\emline{83.00}{34.84}{17}{77.00}{31.83}{18}
\emline{77.00}{31.83}{19}{83.00}{28.82}{20}
\emline{83.00}{28.82}{21}{77.00}{25.81}{22}
\emline{77.00}{25.81}{23}{83.00}{22.80}{24}
\emline{83.00}{22.80}{25}{77.00}{19.78}{26}
\emline{77.00}{19.78}{27}{83.00}{17.20}{28}
\emline{83.00}{17.20}{29}{77.00}{14.19}{30}
\emline{77.00}{14.19}{31}{81.00}{12.04}{32}
\emline{60.00}{49.89}{33}{57.00}{48.17}{34}
\emline{60.00}{49.89}{35}{57.00}{52.04}{36}
\emline{100.00}{49.89}{37}{97.00}{48.17}{38}
\emline{100.00}{49.89}{39}{97.00}{52.04}{40}
\emline{100.00}{12.90}{41}{97.00}{15.05}{42}
\emline{100.00}{9.89}{43}{97.00}{8.17}{44}
\emline{60.00}{9.89}{45}{57.00}{8.17}{46}
\emline{60.00}{12.90}{47}{57.00}{15.05}{48}
\put(1.00,49.89){\makebox(0,0)[lc]{$p_1=(E_1,\vec p)$}}
\put(130.00,49.89){\makebox(0,0)[lc]{$q_1=(E_1,\vec q)$}}
\put(130.00,12.04){\makebox(0,0)[lc]{$q_2=(E_2,-\vec q)$}}
\put(1.00,12.04){\makebox(0,0)[lc]{$p_2=(E_2,-\vec p)$}}
\put(90.00,28.82){\makebox(0,0)[lc]{$\vec k=\vec p-\vec q$}}
\end{picture}
\caption{Feynman diagram of the one-gluon interaction in the quark-diquark
system.}
\end{figure}
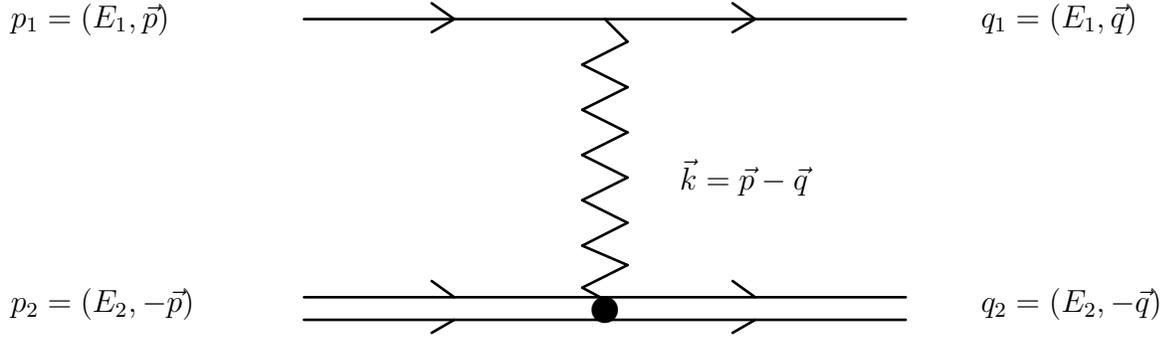
The corresponding expression for the interaction operator has the form:
\begin{equation}
V_{q+SD}^{pert.}(\vec p,\vec q)=\frac{\frac{4}{3}g_s^2}{4\sqrt{\epsilon_1(\vec p)\epsilon_1(\vec q)
\epsilon_2(\vec p)\epsilon_2(\vec q)}}(p_2+q_2)^\mu D_{\mu\nu}(\vec k)\bar u(\vec q)
\gamma^\nu u(\vec p),
\end{equation}
where the interaction vertex of a scalar diquark with a gluon is given
by the factor $ig_s T^a_{ij}(p_2+q_2)^\mu$. Using the standard
representation of the Dirac spinors (6), we get for $V_{q+SD}^{pert.}$
up to terms $O({\vec p}^2/m^2)$ the following expression:
\begin{equation}
V_{q+SD}^{pert.}(\vec p,\vec q)=-\frac{\frac{4}{3}g_s^2 4\pi}{\vec k^2}
\Bigg\{
\Bigg[1+\frac{b^2}{E_1E_2}\Bigg]+\frac{\vec p^2-b^2}{4m_1^2}+\frac{\vec p^2-b^2}{4m_2^2}+
\frac{\vec p^2-b^2}{m_1m_2}-
\end{equation}
\begin{displaymath}
-(\vec p\vec k)\left(\frac{1}{4m_1^2}+\frac{1}{4m_2^2}+
\frac{1}{m_1m_2}\right)
+\frac{\vec k^2}{4m_1m_2}+i\vec S_1[\vec p\times\vec q]\frac{1}
{2m_1}\Bigg(\frac{1}{m_1}+\frac{2}{m_2}\Bigg)\Bigg\}.
\end{displaymath}

Then we add to this expression in the configuration space the
confining potential
\begin{equation}
V_{q+SD}^{nonpert.}(r)=Ar+B,
\end{equation}
which is the same as in the quark-antiquark system. Thus the complete
quasipotential, which is used to calculate the mass spectrum of baryons
with $J^P=\frac{1}{2}^+$ can be represented as follows:
\begin{equation}
V_{q+SD}^{tot.}(r)=Ar+B-\frac{4\alpha_s}{3r}\left(1+\frac{b^2}{E_1E_2}\right)-
\frac{4\pi\alpha_s}{3m_1m_2}\delta(\vec r)+\frac{4\alpha_s}{3m_1m_2}\left(\frac{m_2}{2m_1}
+1\right)\frac{\vec L\vec S_1}{r^3}-
\end{equation}
\begin{displaymath}
-\frac{2\mu\alpha_s^2}{3r^2}\left(\frac{1}{m_1^2}+\frac{1}{m_2^2}+\frac{4}{m_1m_2}\right)
-\frac{4\alpha_s}{3r^3}(\vec r\nabla)\left(\frac{1}{4m_1^2}+\frac{1}{4m_2^2}+\frac{1}
{m_1m_2}\right),
\end{displaymath}
where $m_2$ is the scalar diquark mass, $m_1$ is the light quark
mass and $\mu$ is their reduced mass. The numerical results for the mass spectrum of the ground states
of baryons with $J^P=\frac{1}{2}^+$ on the basis of eqs.~(1) and (13)
are given in Table~2.

\section{Quasipotential for the interaction of a quark with an axial vector
diquark}
As in the previous section, the perturbative part of the interaction
operator for spin $1/2$ and spin $1$ particles is given by the diagram
in Fig.~1. It includes the  matrix element of the chromomagnetic
current operator between states of the axial vector diquark. This matrix element
is defined by three chromomagnetic form factors. Neglecting the
chromoquadrupole moment of the axial vector particle, we can represent the
four vector of the chromomagnetic current in the form \cite{Dv}:
\begin{equation}
j^\mu(\vec p,\vec q)=igT^a_{ij}\xi^\ast\Big\{G_1(k^2)(p_2+q_2)^\mu+\frac{G_2(k^2)}
{m_2}[W^\mu(\vec p_2)(\vec S_2\vec\Delta_2)-(\vec S_2\vec\Delta_2)W^\mu(\vec p_2)]\Big\}\xi,
\end{equation}
\begin{displaymath}
\vec\Delta_2=-\vec q+\frac{\vec p}{m_2}\left(E_2-\frac{\vec q\vec p}{E_2+m_2}
\right),
\end{displaymath}
where $\vec S_2$ is the spin operator of a vector particle,
$W^\mu$ is the four vector of the relativistic spin (the
Pauli-Lubansky vector) with the components:
\begin{equation}
W^0(\vec p_2)=-\vec S_2\vec p,~~~~\vec W(\vec p_2)=m_2\vec
S_2+\frac{\vec p (\vec S_2\vec p)}{E_2+m_2},
\end{equation}
and $\xi$ is the axial vector particle wave function (polarization vector)
in its rest frame. Then the one-gluon exchange quasipotential takes
the form:
\begin{equation} V^{pert.}_{q+AD}(\vec p,\vec
q)=\frac{\frac{4}{3}g^2_s}{4\sqrt{\epsilon_1 (\vec p)\epsilon_1(\vec
q)\epsilon_2(\vec p)\epsilon_2(\vec q)}} D_{\mu\nu}(\vec k)\bar u(\vec
q)\gamma^\nu u(\vec p) \end{equation}
\begin{displaymath}
\times \Big[G_1(k^2)(p_2+q_2)^\mu+G_2(k^2)[\frac{1}{m_2}W^\mu(\vec p_2) (\vec
S_2\vec\Delta_2)-\frac{1}{m_2}(\vec S_2\vec\Delta_2)W^\mu(\vec
p_2)]\Big].
\end{displaymath}
Now we expand eq.~(16) with an accuracy up to terms
$O(\vec p^2/m^2)$. The zeroth and spacial vector components of the current (14)
contain the following terms:
\begin{equation}
\frac{1}{m_2}W^0(\vec p_2)(\vec S_2\vec\Delta_2)-\frac{1}{m_2}(\vec S_2\vec
\Delta_2)W^0(\vec p_2)\cong\frac{i}{m_2}\vec S_2[\vec p\times\vec q],
\end{equation}
\begin{equation}
\frac{1}{m_2}\vec W(\vec p_2)(\vec S_2\vec\Delta_2)-\frac{1}{m_2}(\vec S_2\vec
\Delta_2)\vec W(\vec p_2)\cong{i}[(\vec p-\vec q)\times \vec S_2],
\end{equation}
which produce fine and hyperfine splittings of energy levels. The
chromomagnetic form factor of the axial vector particle $G_2(k^2)$ at
$k^2=0$ defines its total chromomagnetic moment:
$G_2(0)=\mu_V=1+\zeta$. Setting here the
anomalous chromomagnetic moment of the axial vector diquark $\zeta=1$ and
$G_1(0)=1$, we obtain the following expression for
$V_{q+AD}^{pert.}$:

\begin{equation}
V^{pert.}_{q+AD}(\vec p,\vec q)=-\frac{4}{3}\frac{g_s^2~4\pi}{\vec
k^2}\Bigg\{\Bigg(1+\frac{b^2} {E_1E_2}\Bigg)+\frac{\vec
p^2-b^2}{4m_1^2}+\frac{\vec p^2-b^2}{4m_2^2}+ \frac{\vec
p^2-b^2}{m_1m_2}-
\end{equation}
\begin{displaymath}
-(\vec p\vec k)(\frac{1}{4m_1^2}+\frac{1}{4m_2^2}+
\frac{1}{m_1m_2})
+\frac{\vec k^2}{4m_1m_2}+i\vec S_1[\vec p\times\vec q]\frac{1}
{2m_1}\Bigg(\frac{1}{m_1}+\frac{2}{m_2}\Bigg)+
\end{displaymath}
\begin{displaymath}
+i\vec S_2[\vec p\times\vec q]
\frac{1}{2m_2}\Bigg(\frac{1}{m_2}+\frac{2}{m_1}\Bigg)
-\frac{1}{m_1m_2}
[(\vec S_1\vec S_2)\vec k^2-(\vec S_1\vec k)(\vec S_2\vec k)]\Bigg\}.
\end{displaymath}

The nonperturbative part of the interaction potential for a quark and
an axial vector diquark is taken in the form (12). Transforming (19) into
configuration space, we get the complete expression of the
interaction operator for the system $q+AD$:
\begin{equation}
V^{tot.}_{q+AD}(r)=
Ar+B-\frac{4\alpha_s}{3r}\left(1+\frac{b^2}{E_1E_2}\right)-
\frac{4\pi\alpha_s}{3m_1m_2}\delta(\vec r)+\frac{4\alpha_s}{3m_1m_2}\left(\frac{m_2}{2m_1}
+1\right)\frac{\vec L\vec S_1}{r^3}-
\end{equation}
\begin{displaymath}
-\frac{2\mu\alpha_s^2}{3r^2}\left(\frac{1}{m_1^2}+\frac{1}{m_2^2}+\frac{4}{m_1m_2}\right)
-\frac{4\alpha_s}{3r^3}(\vec r\nabla)\left(\frac{1}{4m_1^2}+\frac{1}{4m_2^2}+\frac{1}
{m_1m_2}\right)+
\end{displaymath}
\begin{displaymath}
+\frac{4\alpha_s}{3m_1m_2}\left(\frac{m_1}{2m_2}+1\right)\frac{\vec L\vec S_2}
{r^3}+\frac{32\pi\alpha_s}{9m_1m_2}(\vec S_1\vec S_2)\delta(\vec r)+\frac{4\alpha_s}
{3m_1m_2r^3}[(\vec S_1\vec S_2)-\frac{3}{r^2}(\vec S_1\vec r)(\vec S_2\vec r)].
\end{displaymath}
The potential (20) has been used for calculating the mass spectrum of
baryons with $J^P=\frac{1}{2}^+, \frac{3}{2}^+$ on the basis of the
local quasipotential equation (1). Corresponding numerical results are
presented in Table~2.
\begin{table*}[hbt]
\caption{Masses of baryons (in GeV) containing two heavy quarks.
$[QQ]$ denotes the diquark subsystem with the antisymmetric spin wave
function, $\{QQ\}$ denotes the diquark subsystem with the symmetric
spin wave function ($q = u, d$).}
\begin{center}
\begin{tabular}{|c|c|c|c|c|c|}    \hline
Notation        & Quark    &$J^P$          & $M_B$       & $M_B$    & $M_B$\\
                & content  &               &(our results)&\cite{Ron}&\cite{Korner}     \\  \hline
$\Xi_{cc}$     &$\{cc\}q$     &$\frac{1}{2}^+$&3.66          &3.66     &3.61      \\  \hline
$\Xi_{cc}^\ast$&$\{cc\}q$     &$\frac{3}{2}^+$&3.81          &3.74     &3.68       \\  \hline
$\Omega_{cc}$  &$\{cc\}s$ &$\frac{1}{2}^+$     &3.76          &3.74     &3.71      \\   \hline
$\Omega_{cc}^\ast$&$\{cc\}s$&$\frac{3}{2}^+$     &3.89          &3.82     &3.76       \\   \hline
$\Xi_{bb}$     &$\{bb\}q$     &$\frac{1}{2}^+$&10.23         &10.34    & -     \\  \hline
$\Xi_{bb}^\ast$&$\{bb\}q$     &$\frac{3}{2}^+$&10.28         &10.37    & -      \\  \hline
$\Omega_{bb}$  &$\{bb\}s$     &$\frac{1}{2}^+$     &10.32         &10.37    & -     \\   \hline
$\Omega_{bb}^\ast$&$\{bb\}s$     &$\frac{3}{2}^+$     &10.36         &10.40    & -      \\   \hline
$\Xi_{cb}$     &$\{cb\}q$     &$\frac{1}{2}^+$&6.95          &7.04     & -     \\  \hline
$\Xi'_{cb}$    &$[cb]q$     &$\frac{1}{2}^+$&7.00         &  6.99     & -      \\  \hline
$\Xi_{cb}^\ast$&$\{cb\}q$     &$\frac{3}{2}^+$&7.02         & 7.06     & -      \\  \hline
$\Omega_{cb}$  &$\{cb\}s$     &$\frac{1}{2}^+$&7.05         &7.09   & -     \\  \hline
$\Omega'_{cb}$ &$[cb]s$ &$\frac{1}{2}^+$&7.09         & 7.06   & -      \\  \hline
$\Omega_{cb}^\ast$&$\{cb\}s$  &$\frac{3}{2}^+$&7.11         &7.12   & -      \\  \hline
\end{tabular}
\end{center}
\end{table*}

\section{Conclusions}
In this paper we have carried out the calculation of the mass spectrum of
baryons with two heavy quarks in the quark-diquark approximation on
the basis of a local quasipotential equation. As it follows from
Table~2, our results are in good agreement with calculations made
within other approaches.
One can see from Table 2 that our predictions are especially close to those
obtained in \cite{Ron}. The approach used in \cite{Ron,RD} is based on some
semiempirical regularities of the baryon mass spectrum following from
experimental data and the constituent quark model. The authors of \cite{Ron,RD}
found a successful parametrization of the interpolating curves which describe the
known meson and baryon masses, and give predictions for some yet unobserved mesons
and baryons. Our results for the $B_c$ meson mass spectrum \cite{GMF} are also rather
close to those of \cite{RD}. Our only remark is that one should take into account
the mixing of ${}^3P_1$ and ${}^1P_1$ $B_c$ states. The quoted authors use in particular
the Feynman-Hellmann theorem in order to obtain the monotonic decrease of the
hadron mass spectrum with respect to quark masses. Here they use an oversimplified
assumption about the interquark potential, namely the complete neglection of
the spin-independent part of the Breit-Fermi potential as being unimportant.
This assumption leads finally to some inequalities between quark masses and meson
masses. These inequalities are violated in many constituent quark models (including
ours) in which widely adopted quark masses are used. From our point of view there
is nothing serious in these contradictions and the recovery of the appropriate
spin-independent (flavour-dependent) part of the quark potential \cite{GMF} will
eliminate them. Thus we may conclude that our model correctly describes some important
features of the baryon mass spectrum.

There exists yet another type of baryons with one heavy and two light quarks for
which the quark-diquark approximation gives reasonable results.
Thus, in Ref. \cite{Eb} $Qqq$ baryons were treated by considering the interaction
of the diquark system with the third quark on the basis of additional quark exchange
forces \cite{Cah,Rein,Ebkasch}. The resulting effective lagrangian, which incorporates
heavy quark and chiral symmetry, describes interactions of heavy baryons with
Goldstone bosons in the low energy region. As an interesting application, the
Isgur-Wise form factors have been calculated.

\end{document}